\documentclass{article}
\PassOptionsToPackage{numbers,sort&compress}{natbib}
\usepackage[dblblindworkshop,final]{neurips_2026}
\workshoptitle{Secure and Trustworthy Quantum Machine Learning (SaTQuML)}

\usepackage[utf8]{inputenc}
\usepackage[T1]{fontenc}

\usepackage{amsmath,amssymb}
\usepackage{graphicx}
\usepackage{float}
\usepackage{booktabs}
\usepackage{multirow}
\usepackage{xurl}
\usepackage[hidelinks]{hyperref}
\usepackage{microtype}
\hypersetup{pdftitle={Does the Readout Bypass Leak the Input? A Feature-Visibility Audit of Hybrid Quantum-Classical Models},pdfauthor={Guilin Zhang, Kai Zhao, Xiquan Cui, Henry Heng, Xu Chu, Aletta Johanna Blanken}}

\newcommand{\Qx}{Q(x)}
\newcommand{\xhat}{\hat{x}}

\title{Does the Readout Bypass Leak the Input?\\
A Feature-Visibility Audit of Hybrid Quantum-Classical Models}

\author{%
  \textbf{Guilin Zhang\quad Kai Zhao\quad Xiquan Cui}\\
  \textbf{Henry Heng\quad Xu Chu\quad Aletta Johanna Blanken}\\
  Workday AI Research
}

\begin{document}
\maketitle

\begin{abstract}
Readout-side residual hybrids concatenate raw inputs with measured quantum features. Under single-example gradient sharing, a biased first linear layer admits standard analytic recovery of its input, so the bypass exposes raw coordinates without requiring inversion of the quantum circuit. We audit this mechanism using two tabular datasets, four architectures, and metrics conditioned on feature visibility. Iterative gradient matching gives median full-record PSNR of $73$--$96$\,dB for residual and input-only heads. Quantum-only heads score $8$--$11$\,dB on the full record but $54$--$96$\,dB on the six input coordinates they actually encode. These are reconstruction results for the tested six-input, six-observable circuits, not a general statement about quantum encodings. A loss-threshold membership attack remains near chance. The contribution is a visibility-conditioned privacy audit: omitted coordinates must not be credited as protection supplied by quantum processing, and near-exact PSNR differences must not be interpreted as meaningful privacy rankings. Our findings concern individual gradients and do not establish leakage under aggregation or multiple local training steps.
\end{abstract}

\section{Introduction}
\label{sec:intro}

Quantum machine learning (QML) compresses a classical input into a few measured observables. This narrow readout, the \emph{measurement bottleneck}, can limit downstream accuracy \citep{schuld2020circuit,havlicek2019supervised}. A readout-side residual hybrid measures $\Qx$, concatenates the raw input, $z=[x\,\|\,\Qx]$, and classifies on $z$ \citep{zhang2026readout}. The bypass improves accuracy without increasing quantum depth, which is attractive for federated learning (FL) on resource-constrained clients.

The originating evaluation supports a privacy claim with membership-inference-attack (MIA) AUC near the $0.5$ random baseline \citep{zhang2026readout}. That test asks whether a record was used for training. It does not ask whether an FL server can reconstruct the record from its update. We therefore audit the gradient that the server observes.

The comparison requires care. In our implementation, a $\Qx$-only model reads the first six features, whereas the bypass reads every feature. A full-vector score mixes privacy with feature visibility. Controls with the same MLP widths isolate the raw and quantum paths; total parameter counts differ because their input dimensions differ. We report reconstruction on both the full record and the six shared coordinates. The contribution is this evaluation distinction, not a new gradient-inversion attack.

We connect the known analytic leakage of biased linear layers to this architecture and distinguish direct gradient identifiability from statistical imputation of unseen features.

\begin{figure}[t]
\centering
\includegraphics[width=\textwidth]{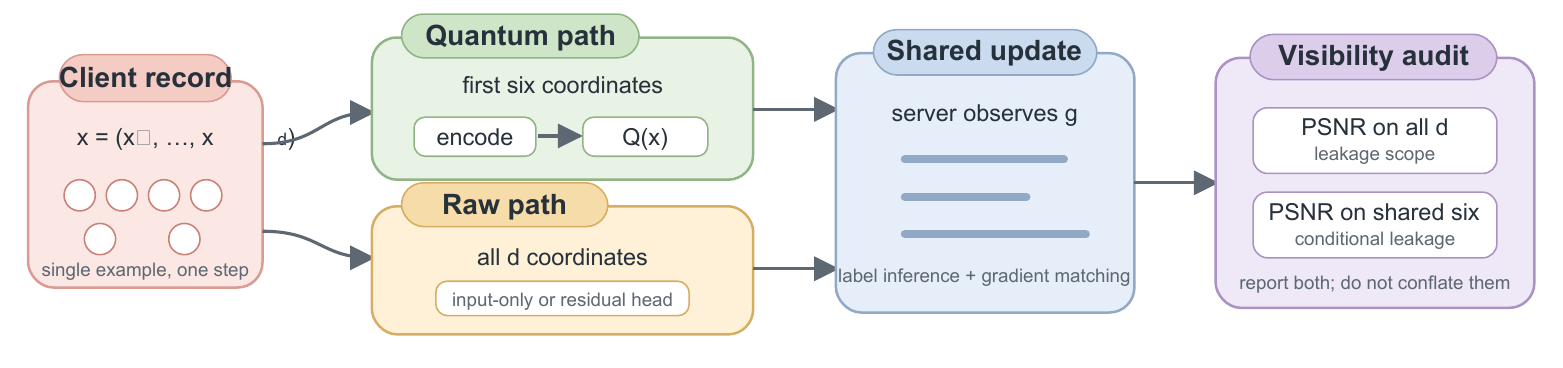}
\caption{Visibility-conditioned audit. Quantum-only heads receive the first six coordinates, while input-only and residual heads receive the full record. We therefore report reconstruction over both the full vector and the six coordinates shared by all quantum models.}
\label{fig:visibility}
\end{figure}

\section{Related work}
\label{sec:related}

\paragraph{Federated quantum learning.}
Federated QML keeps raw data at clients while exchanging hybrid-model updates; some protocols add blind computing and differential privacy \citep{chen2021federated,li2021blind}. Local records alone do not make those updates private. We audit this distinction for a readout-side residual architecture.

\paragraph{Privacy of shared updates.}
FL updates can leak membership and unintended features \citep{nasr2019comprehensive,melis2019exploiting,carlini2022membership}. Gradient reconstruction targets the record itself \citep{zhu2019deep,geiping2020inverting,balunovic2022bayesian}; malicious servers can amplify it even for large batches \citep{fowl2022robbing,wen2022fishing,boenisch2023curious}. Secure aggregation hides individual updates from honest aggregators, but repeated observations or inconsistent models weaken this guarantee \citep{bonawitz2017practical,lam2021gradient,pasquini2022eluding}. We isolate one unaggregated, single-example update received by an honest-but-curious server.

\section{Audit methodology}
\label{sec:threat}

\paragraph{The architecture under audit.}
Our implementation of the readout-side design \citep{zhang2026readout} maps the first six features to six Pauli-$Z$ expectations $\Qx$. The residual forms $z=[x\,\|\,\Qx]$ and applies a biased linear projection followed by an MLP. The pure-quantum model uses a biased linear classifier on $\Qx$; the other quantum-only head uses the same MLP widths as the residual. The first-six selection is our explicit implementation choice. The original paper does not fully specify how higher-dimensional inputs map to six encoding angles; our visibility result therefore does not establish that its entire pipeline discards those same coordinates.

\paragraph{Federated setting and adversary.}
The honest-but-curious server knows trained weights and receives all parameter gradients, including the first weight and bias gradients, for one record. This models direct gradient sharing or one plain SGD step with known learning rate, from which $g=(\theta-\theta')/\eta$ is recoverable. Training uses Adam minibatches of eight; the attack separately computes a single-record diagnostic gradient. It does not observe an actual multi-step FedAvg update \citep{mcmahan2017communication}. No raw coordinates are sent explicitly.

\paragraph{Analytic baseline.}
For a first affine layer $a=Wz+b$, write $g_W=\partial\mathcal L/\partial W$ and $g_b=\partial\mathcal L/\partial b$. For any row $j$ with $(g_b)_j\ne0$,
\begin{equation}
g_{W,j:}=(g_b)_jz^\top,\qquad z^\top=g_{W,j:}/(g_b)_j.
\label{eq:analytic}
\end{equation}
This standard leakage mechanism \citep{phong2018privacy,geiping2020inverting} recovers $x$ for raw-input heads and $Q(x)$ for quantum-only heads. Recovering $Q(x)$ does not by itself invert the circuit. Appendix~\ref{app:analytic} states the conditions and numerical limitations.

\paragraph{Reconstruction objective.}
For comparison, the existing iterative attack matches $G(u,c)=\nabla_\theta\mathcal{L}(f_\theta(u),c)$. iDLG infers $\hat y$ from the most negative final bias gradient \citep{zhao2020idlg}. We solve \citep{geiping2020inverting,zhu2019deep,yin2021see}
\begin{equation}
\xhat=\arg\min_{u\in[0,1]^d}\left[1-
\frac{\langle G(u,\hat y),g\rangle}{\|G(u,\hat y)\|_2\,\|g\|_2}\right].
\label{eq:attack}
\end{equation}
The box constraint matches min--max preprocessing. We omit total variation because adjacent tabular columns have no spatial meaning.

\paragraph{Visibility-conditioned metrics and controls.}
For each metric $m$, we report $m(x,\xhat)$ on all $d$ coordinates and $m(x_S,\xhat_S)$ on $S=\{1,\ldots,6\}$. Shared-six scores remove the contribution of omitted coordinates but do not make the full architectures visibility-matched: the residual still receives more inputs. Appendix~\ref{app:threshold} also reports success at explicit MSE thresholds.

\paragraph{Why membership is insufficient here.}
MIA near $0.5$ indicates that membership is hard to distinguish from model outputs. It does not measure reconstruction from a training update. The original evaluation names gradient inversion and PSNR among its metrics but reports MIA AUC and an MSE privacy score without a PSNR result \citep{zhang2026readout}. We supply the reconstruction audit and retain a simple loss-threshold MIA as a diagnostic comparison.

\section{Experiments}
\label{sec:exp}

\paragraph{Setup.}
We use Wine and Breast Cancer with one fixed stratified 75/25 split (seed 42) and training-only min--max scaling. The six-qubit map encodes the first six features and uses two $R_Y/R_Z$ and CZ layers. We compare pure quantum, $Q(x)$-only, input-only, and residual heads. The latter three have the same projection and MLP widths, not identical parameter counts. Accuracy is measured on held-out data; attacks target training records.

\paragraph{Attack.}
We use cosine gradient matching \citep{geiping2020inverting} with $\xhat\in[0,1]$. Because the features have no spatial order, we remove the image-style total-variation prior. The class is inferred from the final bias gradient. For each architecture we train three seeds and attack five records per seed with two random restarts and 200 optimization steps, yielding 15 attacked records per dataset. We report median and interquartile range (IQR) for MSE, PSNR (peak 1.0), and cosine similarity over all features and over the first six shared features. For comparison, the MIA score is the ROC AUC obtained by thresholding negative per-example loss on the train and test partitions.

\paragraph{Results.}
Table~\ref{tab:main} reports iterative reconstruction. Residual and input-only full-vector medians are $73$--$96$\,dB, consistent with the analytic vulnerability. Quantum-only medians are $8$--$11$\,dB on the full vector but $54$--$96$\,dB on the shared six. All attacks infer the label correctly. Omitted coordinates are not directly identifiable from these gradients, although feature correlations could support imputation. We did not evaluate that additional inference.

\begin{table}[t]
\caption{Held-out accuracy and privacy audit. PSNR is the median over 15 attacked training records; higher means easier reconstruction. ``Shared'' evaluates the first six coordinates visible to every quantum model. Accuracy and MIA AUC are means over three seeds.}
\label{tab:main}
\centering
\small
\begin{tabular}{llcccc}
\toprule
Dataset & Architecture & Test acc. & MIA AUC & PSNR all & PSNR shared \\
\midrule
\multirow{4}{*}{Wine}
 & Pure quantum        & .896 & .477 & 11.08 & 95.56 \\
 & $Q(x)$-only MLP     & .881 & .527 & 10.55 & 67.90 \\
 & Input-only          & 1.000 & .512 & 85.77 & 88.77 \\
 & Residual bypass     & .978 & .491 & 72.95 & 75.25 \\
\midrule
\multirow{4}{*}{Breast Cancer}
 & Pure quantum        & .909 & .540 & 8.06 & 90.28 \\
 & $Q(x)$-only MLP     & .881 & .530 & 8.20 & 53.95 \\
 & Input-only          & .967 & .515 & 96.39 & 95.23 \\
 & Residual bypass     & .967 & .517 & 94.96 & 93.57 \\
\bottomrule
\end{tabular}
\end{table}

\paragraph{Reading the numbers.}
Near-exact PSNR differences reflect optimization and numerical residuals, not meaningful privacy rankings. The raw-feature path suffices for exposure by Equation~\ref{eq:analytic}; quantum interaction is unnecessary. The residual's low PSNR quartiles show that 200-step matching can fail to exploit an analytically vulnerable layer. Input-only accuracy also matches or exceeds residual accuracy on these datasets. MIA AUC remains $.477$--$.540$ despite reconstruction, illustrating that membership and reconstruction measure different properties.

\section{Discussion}
\label{sec:disc}

\paragraph{What the bypass changes.}
The bypass expands the coordinates exposed to standard affine-layer leakage. A $Q(x)$-only gradient has no direct dependence on omitted coordinates; this does not preclude statistical inference using correlated visible features. Recoverability of six encoded inputs from six observables and circuit gradients in this experiment does not establish invertibility of compressive quantum encodings.

\paragraph{Membership and reconstruction are not interchangeable.}
MIA AUC near $0.5$ indicates weak membership evidence under a particular attack \citep{shokri2017membership}. AUC is itself an average-case summary and should not be read as a privacy certificate \citep{carlini2022membership}. More importantly here, it says little about whether a specific client input is recoverable from a shared gradient. Quantum encoding is sometimes credited with privacy from state collapse \citep{farokhi2024maximal}; our shared-feature result cautions against transferring that intuition to differentiable training gradients.

\paragraph{Defenses and limitations.}
DP-SGD \citep{abadi2016deep}, secure aggregation \citep{bonawitz2017practical}, and larger batches \citep{huang2021evaluating} change exposure. Our evidence covers two small datasets, one split, 15 record-seed attacks per architecture, and noiseless simulation. We have not measured the analytic baseline's empirical success rate, a classical six-input/six-output control, compressive or all-feature encodings, imputation, larger batches, multiple local steps, or withholding the first-layer gradients. These remain necessary for broader claims.

\paragraph{Implications for evaluation.}
A defensible privacy table should state, for every architecture, which input coordinates affect the transmitted update and what aggregation precedes server access. A full-vector error is comparable only when architectures observe the same coordinates. Where visibility differs, shared-coordinate and full-record metrics should appear side by side. MIA should remain a separate evaluation and, when central to a claim, should go beyond aggregate AUC. These reporting choices are not a defense; they prevent a missing input path from being mislabeled as privacy.

\section{Conclusion}
\label{sec:conclusion}

A readout bypass exposes raw features to known affine-layer gradient leakage under the stated conditions. In our six-input circuits, visible features are also reconstructable by iterative matching. Audits should specify the server's observable, condition metrics on input visibility, and distinguish reconstruction from membership inference.

\clearpage
\bibliographystyle{plainnat}
\bibliography{references}

\clearpage
\appendix

\section{Reproducibility details}
\label{app:repro}

\paragraph{Data and preprocessing.}
Wine contains 178 examples, 13 features, and 3 classes; Breast Cancer contains 569 examples, 30 features, and 2 classes. The audit script constructs one stratified 75/25 split using the first seed (42) and fits min--max scaling on its training partition. All model seeds $\{42,43,44\}$ reuse that split. Each seed selects five training records without replacement. The saved indices confirm 15 distinct attacked records per dataset and architecture, but these are not independent data splits; architectures use the same selected indices. No test record is used for training, scaling, or reconstruction. Breast Cancer uses all 30 raw features in our audit, rather than the PCA-reduced representation described in the originating paper.

\paragraph{Models and optimization.}
The quantum feature map applies $R_Y(\pi\tanh x_j)$ to the first six coordinates, a CNOT ladder, two variational layers of per-qubit $R_Y/R_Z$ rotations and a CZ ring, then six Pauli-$Z$ measurements. The projection head is $\mathrm{Linear}(p,16)$--ReLU--$\mathrm{Linear}(16,16)$--ReLU--$\mathrm{Linear}(16,k)$; every linear layer includes a bias. ``Matched'' in the original control name refers only to hidden widths and depth, not total parameters (Table~\ref{tab:accmia}). Models train for 30 epochs with Adam, learning rate $.01$, cross-entropy loss, and minibatches of 8. Quantum differentiation uses PennyLane's noiseless \texttt{default.qubit} simulator with backpropagation. Attacks use the trained model and all trainable parameter gradients.

\paragraph{Attack and statistics.}
For every model seed we attack five records, each with two random restarts, 200 Adam steps, and learning rate $.05$. The best cosine-gradient objective is retained. Inputs are clipped to $[0,1]$ after every step. Labels are not supplied to the attacker; they are inferred from the final bias gradient and are correct in all 120 attacks. We report medians and IQRs across the 15 attacked records per dataset and architecture. The diagnostic MIA assigns score $-\mathcal L(f_\theta(x),y)$ and computes ROC AUC across all train and test records, separately for each seed; Table~\ref{tab:accmia} reports mean and standard deviation across seeds.

\begin{table}[H]
\caption{Capacity, held-out accuracy, and loss-threshold MIA AUC (mean $\pm$ standard deviation over three seeds).}
\label{tab:accmia}
\centering
\small
\begin{tabular}{llrrr}
\toprule
Dataset & Architecture & Parameters & Test accuracy & MIA AUC \\
\midrule
\multirow{4}{*}{Wine}
 & Pure quantum        & 45  & $.896\pm.010$ & $.477\pm.011$ \\
 & $Q(x)$-only MLP & 459 & $.881\pm.021$ & $.527\pm.012$ \\
 & Input-only          & 547 & $1.000\pm.000$ & $.512\pm.003$ \\
 & Residual bypass     & 667 & $.978\pm.000$ & $.491\pm.016$ \\
\midrule
\multirow{4}{*}{Breast Cancer}
 & Pure quantum        & 38  & $.909\pm.006$ & $.540\pm.003$ \\
 & $Q(x)$-only MLP & 442 & $.881\pm.032$ & $.530\pm.005$ \\
 & Input-only          & 802 & $.967\pm.012$ & $.515\pm.001$ \\
 & Residual bypass     & 922 & $.967\pm.009$ & $.517\pm.006$ \\
\bottomrule
\end{tabular}
\end{table}

\section{Complete reconstruction results}
\label{app:results}

Table~\ref{tab:fullmetrics} gives full-vector medians; Table~\ref{tab:sharedmetrics} evaluates the first six coordinates. The full-vector score includes omitted coordinates that the iterative attack leaves unconstrained. It does not measure the best achievable reconstruction when an adversary can exploit feature correlations. Very high PSNR values should be read as near-exact reconstruction, not a calibrated ordering of privacy.

\begin{table}[H]
\caption{Full-vector reconstruction medians over 15 records. Higher PSNR and cosine, and lower MSE, indicate greater exposure.}
\label{tab:fullmetrics}
\centering
\small
\begin{tabular}{llrrr}
\toprule
Dataset & Architecture & MSE & PSNR (dB) & Cosine \\
\midrule
\multirow{4}{*}{Wine}
 & Pure quantum        & $7.80\mathrm{e}{-2}$ & 11.08 & .8608 \\
 & $Q(x)$-only MLP & $8.81\mathrm{e}{-2}$ & 10.55 & .8314 \\
 & Input-only          & $2.65\mathrm{e}{-9}$ & 85.77 & 1.0000 \\
 & Residual bypass     & $5.07\mathrm{e}{-8}$ & 72.95 & 1.0000 \\
\midrule
\multirow{4}{*}{Breast Cancer}
 & Pure quantum        & $1.56\mathrm{e}{-1}$ & 8.06 & .7243 \\
 & $Q(x)$-only MLP & $1.51\mathrm{e}{-1}$ & 8.20 & .7354 \\
 & Input-only          & $2.30\mathrm{e}{-10}$ & 96.39 & 1.0000 \\
 & Residual bypass     & $3.19\mathrm{e}{-10}$ & 94.96 & 1.0000 \\
\bottomrule
\end{tabular}
\end{table}

\begin{table}[H]
\caption{Shared-six reconstruction medians over 15 records.}
\label{tab:sharedmetrics}
\centering
\small
\begin{tabular}{llrrr}
\toprule
Dataset & Architecture & MSE & PSNR (dB) & Cosine \\
\midrule
\multirow{4}{*}{Wine}
 & Pure quantum        & $2.78\mathrm{e}{-10}$ & 95.56 & 1.0000 \\
 & $Q(x)$-only MLP & $1.62\mathrm{e}{-7}$ & 67.90 & 1.0000 \\
 & Input-only          & $1.33\mathrm{e}{-9}$ & 88.77 & 1.0000 \\
 & Residual bypass     & $2.99\mathrm{e}{-8}$ & 75.25 & 1.0000 \\
\midrule
\multirow{4}{*}{Breast Cancer}
 & Pure quantum        & $9.37\mathrm{e}{-10}$ & 90.28 & 1.0000 \\
 & $Q(x)$-only MLP & $4.03\mathrm{e}{-6}$ & 53.95 & 1.0000 \\
 & Input-only          & $3.00\mathrm{e}{-10}$ & 95.23 & 1.0000 \\
 & Residual bypass     & $4.40\mathrm{e}{-10}$ & 93.57 & 1.0000 \\
\bottomrule
\end{tabular}
\end{table}

\begin{table}[H]
\caption{PSNR IQR over 15 attacked records. Intervals are in dB and expose the broad record-to-record variability hidden by a single median.}
\label{tab:iqr}
\centering
\small
\begin{tabular}{llrr}
\toprule
Dataset & Architecture & All features & Shared six \\
\midrule
\multirow{4}{*}{Wine}
 & Pure quantum        & 10.1--12.8 & 87.3--100.3 \\
 & $Q(x)$-only MLP & 10.1--11.5 & 32.7--92.8 \\
 & Input-only          & 62.8--96.9 & 63.2--97.8 \\
 & Residual bypass     & 42.9--93.5 & 45.5--95.4 \\
\midrule
\multirow{4}{*}{Breast Cancer}
 & Pure quantum        & 7.0--8.9 & 30.8--101.7 \\
 & $Q(x)$-only MLP & 7.4--9.0 & 17.8--91.5 \\
 & Input-only          & 81.8--102.1 & 87.2--102.2 \\
 & Residual bypass     & 66.5--99.3 & 67.9--100.2 \\
\bottomrule
\end{tabular}
\end{table}

\clearpage
\section{Analytic leakage and numerical recovery criteria}
\label{app:analytic}

For one example, the chain rule gives $g_W=\delta z^\top$ and $g_b=\delta$ at a first affine layer, regardless of the downstream differentiable network. A row with nonzero $\delta_j$ yields Equation~\ref{eq:analytic}. In finite precision, selecting the row with the largest $|(g_b)_j|$ avoids dividing by an unnecessarily small signal. An alternative combines rows as $z=g_W^\top g_b/\|g_b\|_2^2$ when the denominator is nonzero. Neither formula requires the example's class label. These are applications of established gradient leakage, not new attacks \citep{phong2018privacy,geiping2020inverting}.

For the residual, $z=[x\|Q(x)]$, so the first $d$ recovered coordinates are the raw input. The input-only head directly yields $z=x$. For either quantum-only head the formula yields $z=Q(x)$; recovering $x$ still requires inversion or additional information from circuit-parameter gradients. Six measured expectations need not define an injective map, even with six inputs. Our optimization results establish recovery for the tested records and parameters only.

The identity requires individual weight and bias gradients and a nonzero backpropagated signal. With zero signals, missing bias gradients, gradient perturbation, or aggregation, it does not give the same recovery guarantee. For a batch, $g_W=\sum_i\delta_i z_i^\top/B$ and $g_b=\sum_i\delta_i/B$; their row ratio is generally a weighted mixture rather than any one record. Likewise, a multi-step Adam or FedAvg parameter delta is not the single gradient used here. The minibatch training procedure therefore does not itself show that an ordinary federated server receives the diagnostic gradients we attack.

The analytic result is exact under its algebraic assumptions. We have not run a separate empirical gradient-ratio evaluation on saved checkpoints, so no observed analytic success rate is claimed. The iterative attack's finite budget can leave avoidable residual error; its lower quartiles do not establish protection against the analytic attack.

\subsection{Thresholded recovery from saved iterative attacks}
\label{app:threshold}
Table~\ref{tab:threshold} summarizes the original per-target records at MSE thresholds $10^{-6}$ and $10^{-8}$ (RMSE $10^{-3}$ and $10^{-4}$ for features scaled to $[0,1]$). These are explicitly defined numerical tolerances, not claims of bitwise-exact recovery. Rates use the same 15 record-seed pairs as the PSNR summaries. They quantify the existing iterative attack, not the unmeasured analytic baseline. The implementation floors MSE at $10^{-12}$ when computing PSNR, corresponding to a 120\,dB ceiling.
\begin{table}[H]
\caption{Iterative recovery counts at explicit MSE thresholds, recomputed from saved per-target results. Each denominator is 15 record-seed pairs.}
\label{tab:threshold}
\centering
\small
\begin{tabular}{llcccc}
\toprule
 & & \multicolumn{2}{c}{All coordinates} & \multicolumn{2}{c}{Shared six} \\
Dataset & Architecture & $\leq10^{-6}$ & $\leq10^{-8}$ & $\leq10^{-6}$ & $\leq10^{-8}$ \\
\midrule
Wine & Pure quantum & 0/15 & 0/15 & 14/15 & 14/15 \\
Wine & $Q(x)$-only MLP & 0/15 & 0/15 & 9/15 & 7/15 \\
Wine & Input-only & 11/15 & 10/15 & 11/15 & 10/15 \\
Wine & Residual bypass & 9/15 & 7/15 & 9/15 & 6/15 \\
\midrule
Breast Cancer & Pure quantum & 0/15 & 0/15 & 10/15 & 10/15 \\
Breast Cancer & $Q(x)$-only MLP & 0/15 & 0/15 & 7/15 & 7/15 \\
Breast Cancer & Input-only & 13/15 & 12/15 & 13/15 & 12/15 \\
Breast Cancer & Residual bypass & 11/15 & 11/15 & 11/15 & 11/15 \\
\bottomrule
\end{tabular}
\end{table}

\end{document}